\documentclass[twocolumn,showpacs,preprintnumbers,amsmath,amssymb,a4paper]{revtex4}
\usepackage{amsmath}
\usepackage{amsfonts}
\usepackage{amsthm}
\usepackage{amssymb}
\usepackage{graphicx}
\usepackage{mathrsfs}
\usepackage{latexsym}
\usepackage{graphics}

\newcommand{\be}{\begin{equation}}
\newcommand{\ee}{\end{equation}}

\begin{document}

\title{Confined Shocks inside Isolated Liquid Volumes -- A New Path of Erosion?}
\author{D. Obreschkow$^1$}
\author{N. Dorsaz$^2$}
\author{P. Kobel$^1$}
\author{A. de Bosset$^1$}
\author{M. Tinguely$^1$}
\author{J. Field$^4$}
\author{M. Farhat$^1$}

\affiliation{
$^1\,$EPFL, Laboratoire des Machines Hydrauliques, Av. Cour 33bis, 1007 Lausanne, Switzerland\\
$^2\,$Department of Chemistry, University of Cambridge, Cambridge CB2 1EW, UK\\
$^3\,$Physics and Chemistry of Solids, Cavendish Laboratory, Cambridge CB3 OHE, UK}


\date{\today}

\begin{abstract}
The unique confinement of shock waves inside \emph{isolated} liquid volumes amplifies the density of shock--liquid interactions. We investigate this universal principle through an interdisciplinary study of shock-induced cavitation inside liquid volumes, isolated in 2 and 3 dimensions. By combining high-speed visualizations of ideal water drops realized in microgravity with smoothed particle simulations we evidence strong shock-induced cavitation at the focus of the confined shocks. We extend this analysis to ground-observations of jets and drops using an analytic model, and argue that cavitation caused by trapped shocks offers a distinct mechanism of erosion in high-speed impacts ($\gtrsim100\rm\,m\,s^{-1}$).
\end{abstract}

\maketitle

\begin{figure}[t!]
  \includegraphics[width=8.3cm]{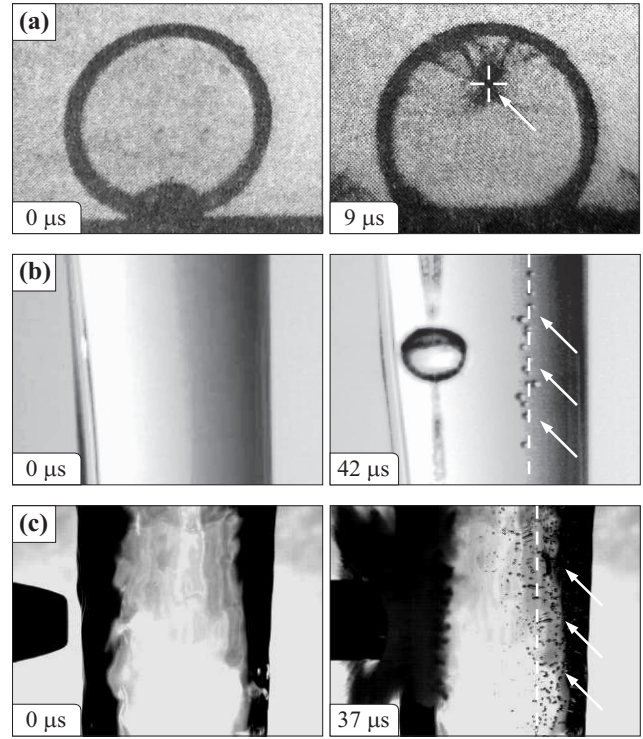}
	  \caption{Examples of cavitation (arrows) produced by reflected shock waves. (a) Liquid drop (88\% water, 12\% gelatine, diameter $D=10\rm\,mm$) impacting on a hard solid surface at $v=\rm110\,m\,s^{-1}$ \citep[reprint from Ref.~][]{Field1989}; (b) Water jet ($D=6\rm\,mm$) shocked by a laser pulse ($\rm energy=30\,mJ$) on the left side \citep[setup explained in Ref.~][]{Robert2007}; (c) Water jet ($D=22\rm\,mm$) impacted by 9\,mm-projectile at $v=\rm200\,m\,s^{-1}$ [collaboration with the Swiss army]. Dashed lines indicate the position, where the cavitation should occur according to eq.~(\ref{eq_mu}), corrected for optical refraction \citep{Kobel2009}.}
  \label{fig_universality}
\end{figure}

Shock waves in liquids are a common cause of cavitation \citep{Kedrinskii2007,Tomita1991,Hagenson1998,Wolfrum2003,Sankin2005}, in particular when shocks are reflected and focussed \citep{Lindau2003}. Famous examples include shock-induced cavitation in lithotripsy \citep{Tomita1994,Leighton2010} and the `white crown' on the sea surface following an underwater detonation \citep{Wilson1947,Chapman1965,Wentzell1969}. However, little is known about shock-induced cavitation inside ``isolated'' liquid volumes \citep{Heijnen2009}, which are completely bounded by a free surface. The crucial feature of such systems is their unique confinement: the closed surface acts as a mirror trapping the shock and amplifying its local interaction with the fluid. Experimental hints for the importance of this mechanism for generating cavitation were provided by some of the earliest high-speed visualizations of shocked drops \citep[Ref.][see reprint in Fig.~\ref{fig_universality}a]{Field1989}.

In this letter, we study the amplification of shock-induced cavitation in liquid volumes isolated in three dimensions (3-d), such as drops, and in two dimensions (2-d), such as jets. Illustrations of shock-driven cavitation in both cases are provided in Fig.~\ref{fig_universality}. To understand this cavitation, we perform a systematic experimental study of shock-induced cavitation inside large, spherical water drops. These drops are realized in microgravity conditions aboard parabolic flights (European Space Agency, 42nd Parabolic Flight Campaign). In parallel to those experiments, we provide a quantitative explanation of the observed cavitation patterns through numerical simulations of dissipative shocks inside spheres, and derive an analytic model to predict the location of shock-induced cavitation. Thereby we demonstrate that shock-induced cavitation in isolated volumes is a universal phenomenon. Finally, we discuss a potential implication of cavitation caused by trapped shocks for drop erosion of solid surfaces.

\begin{figure}[t]
  \includegraphics[width=8.3cm]{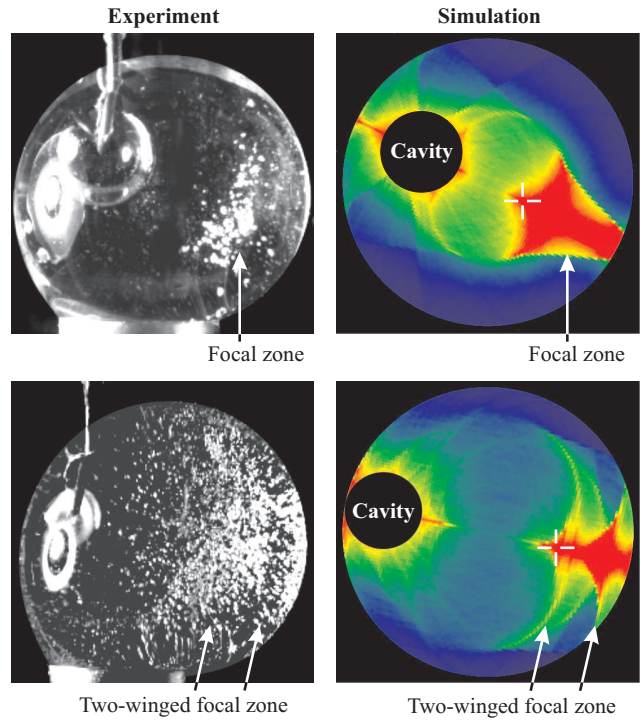}
  \caption{Spherical water drop ($D=\rm22\,mm$) $80\,\mu\rm s$ after the generation of a spherical shock at the location labelled ``cavity'' (since here the primary cavitation bubble forms in the case of a plasma-driven shock). Left images show the observed shock-induced cavitation, while the right images are simulated counter-parts; the colors range from lowest (blue) to highest (red) energy density. The eccentricities of the shock origins are $\epsilon=0.4$ (upper) and $\epsilon=0.6$ (lower), though they appear to be larger due to optical refraction \citep[refraction model in Ref.][]{Kobel2009}. Dashed crosses indicate where the strongest cavitation should occur according to eq.~(\ref{eq_mu}).}
  \label{fig_comparison}
\end{figure}

Our microgravity experiment can be seen as an ideal laboratory for the study of shock dynamics inside stable drops. The setup \citep[details in Ref.][]{Kobel2009} can produce a spherical drop of demineralized water (diameter $D=16-26\rm\,mm$). This drop is smoothly expelled through an injector tube, which also serves as a permanent attach-point for the drop (see Fig.~\ref{fig_comparison}, left). A movable pair of electrodes penetrating the drop from the top in Fig.~\ref{fig_comparison} releases a fast ($10\rm\,ns$) discharge at a specific location within the drop. This discharge forms a supersonically expanding point-plasma \citep[physics in Ref.][]{Kennedy1997}, which generates a spherical shock wave and a vapor bubble \citep{Obreschkow2006}, called the ``primary cavitation bubble''. We here focus on the shock wave, while regarding the primary cavitation bubble as a welcome side-effect to estimate the energy of the shock wave. In fact, early studies of laser-induced point-plasmas \citep{Vogel1996,Kobel2009} evidenced that the shock energy $E_{\rm s}$ approximately equals the bubble energy $E_{\rm b}=4\pi/3\cdot r_{\rm max}^3\cdot(p_\infty-p_v)$, where $r_{\rm max}$ is the maximal bubble radius, $p_\infty=80\rm\,kPa$ is the static water pressure (= ambient pressure in the aircraft cabin), and $p_{\rm v}=3.2\rm\,kPa$ is the water vapour pressure at the working temperature $T\approx\rm25\,^\circ C$. Here we study cases in the range $r_{\rm max}=2-4\rm\,mm$, hence $E_{\rm s}\approx E_{\rm b}\approx3-20\rm\,mJ$. The fast phenomena produced by the shock are recorded using a high-speed camera (Photron Ultima APX) at up to $120\,000$ frames/sec and $50\rm\mu m$ spatial resolution.

The data is acquired in 36 microgravity cycles, each of $\rm20\,s$ duration. Each cycle allows us to generate a single water drop and to release one shock wave within this drop. The free parameters defining such a cycle are the drop diameter $D$, the shock energy $E_{\rm s}$, and the location of the shock center. The latter is expressed by the eccentricity $\epsilon\equiv s/(D/2)\in[0,1]$, where $s$ is the distance between the drop center and the shock center. The results presented in this letter rely on the 18 parameter configurations corresponding to all possible combinations of $D\rm\in\{16\,mm, 22\,mm\}$, $E_s\rm\in\{3\,mJ,10\,mJ,20\,mJ\}$, and $\epsilon\in\{0.2,0.4,0.6\}$. Each configuration is repeated twice to verify the repeatability of the data.

All 36 high-speed visualizations exhibit the same dominant feature: About $\rm10\,\mu s$ after the shock generation thousands of submillimetric bubbles appear synchronously in the hemisphere opposite the origin of the shock (Fig.~\ref{fig_comparison} left). The fastest visualizations ($120\,000$ frames/sec) uncover that these micro-bubbles grow and collapse, their longest life-times lying around $50\,\mu s$. This transient behavior and the coherent formation of the bubbles at the instant of the shock transition disclose that the bubbles are a form of shock wave-induced cavitation \citep{Kedrinskii2007,Tomita1991,Hagenson1998,Wolfrum2003,Arora2007,Leighton1989} -- a phenomenon known as ``secondary cavitation'' in other situations. The largest bubbles have diameters of $d\approx\rm0.5mm$, which in the Rayleigh-model \citep{Rayleigh1917} of spherical cavitation bubbles implies life-times of $t_{\rm R}=0.915\,d\,[\rho/(p_\infty-p_v)]^{1/2}\approx\rm50\,\mu s$, consistent with our observations.

The discrepancy between the high bubble-density seen in Fig.~\ref{fig_comparison} (left) and the faint traces detected by others \citep{Tomita1991} can be explained by the particular confinement of our shock. The free drop surface causes an elastic reflection of a pressure wave; hence the shock bounces back and forth while successively dissipating its energy to shock-induced cavities. To check if the full shock energy is converted into cavitation bubbles, we estimate the total volume $V$ of these bubbles from their radii (corrected for optical refraction by using our optical model in Ref.~\citealp{Kobel2009}). The implied energy $E=V\cdot(p_\infty-p_v)$ is systematically consistent (within $20\%$ measurement uncertainties) with the original shock energy $E_{\rm s}$.

What is the role of gas and impurities contained in the water? Shock-induced cavitation bubbles arise when the shock pressure excites pre-existing nuclei \citep{Kedrinskii2007}. A priori, one could therefore expect that the amount of shock-induced cavitation inside a shocked drop depends on the initial nuclei content. However, in the particular case where the shock keeps bouncing off the free surface until its entire energy is converted into bubbles, the nuclei content cannot play a major role for the amount of shock-induced cavitation. Our observations of strong shock-induced cavitation in very clean water confirm this peculiar feature of confined shocks.

Depending on the experimental parameters, the cloud of shock-induced bubbles appears in different intensities and geometries (Fig.~\ref{fig_comparison}). A systematic investigation of all samples uncovers that the intensity (size and number) of bubbles varies with the shock energy $E_{\rm s}$ and drop diameter $D$. In return, the geometry of the bubble-cloud exclusively depends on the eccentricity $\epsilon$, as will be explained hereafter.


To understand the patterns formed by shock-induced cavitation bubbles, we implement a new simulation scheme of dissipative shocks inside spherical drops based on a smoothed particle approach. In our model, the shock is considered as a spherical shell that isotropically expands from a single point, specified by its eccentricity $\epsilon$. This shell is represented by $N=10^5$ particles, which initially propagate on straight lines at the speed of sound $c=1500\rm\,m\,s^{-1}$. When reaching the drop surface they are reflected elastically. Each particle carries a shock energy $E_{\rm p}(t)\propto(\delta p)^2$, where $\delta p$ is the pressure fluctuation relative to the equilibrium state. (Note that $\delta p$ changes its sign when reflecting at the free surface, while the energy remains unchanged.) As these ``shock quanta'' propagate across the liquid they form cavitation via evaporation in the tensile case ($\delta p<0$), or through forcing the collapse and subsequent rebound of existing nuclei in the compressive case ($\delta p>0$). In both cases, the generation of cavitation bubbles decreases the energy of the shock by decreasing $|\delta p|$. We assume that this decrease in energy happens at a constant fraction per unit time (independently of the sign of $\delta p$), as typical for most dissipative processes. Thus, $E_{\rm p}(t)=E_{\rm s}/N\cdot\exp(-t/\tau)$, where $\tau$ is the time, in which the shock dissipates $(1-1/e)\approx63\%$ of its energy. Here we choose $\tau=20\rm\,\mu s$ since the shock-induced bubbles in Fig.~\ref{fig_comparison} (left) reach an integrated potential energy, measured through the bubble sizes, of $(0.6\pm0.2)E_{\rm s}$ within $20\rm\,\mu s$. The simulation progresses at a discrete time step ${\rm d}t=10^{-8}\rm\,s$, chosen sufficiently small to obtain converging results. In each step and for each particle, the dissipated energy $|{\rm d}E_{\rm p}(t)|=-\dot{E}_{\rm p}\,{\rm d}t$ is transcribed to the liquid at the location of the particle using a 3-d Gaussian smoothing kernel with a variance equal to the mean separation between neighboring particles. To tackle the energy acquired at different locations in the drop, the latter is discretized on a regular mesh of $256^3\approx1.7\cdot10^7$ cubic cells.

The output of these numerical simulations is a 3-d density map of the energy dissipated by a spherical shock trapped inside a spherical drop. To compare these maps against the 2-d images of shock-induced cavitation inside drops, the simulated maps were projected onto a plane, while accounting for the optical refraction at the water surface \citep[according to our model in Ref.][]{Kobel2009}. Fig.~\ref{fig_comparison} compares the simulation against the observations for the two eccentricities $\epsilon=0.4, 0.6$. The patterns of the simulated energy density and the observed shock-induced cavitation exhibit remarkable similarities: 
in both cases (i) the highest densities appear opposite the origin of the shock wave at a comparable eccentricity, and (ii) for high eccentricities ($\epsilon>0.5$) the bubble-patterns spread out in two wings. This match between simulation and experiment confirms that the patterns of shock-induced cavitation can be understood in terms of reflected dissipative waves.


\begin{figure}[t]
  \includegraphics{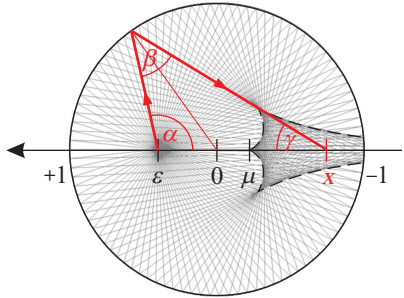}
  \caption{Circular wave reflected inside a circle. The thick line highlights a selected path. The ensemble of all reflected paths defines the dashed envelope, called ``catacaustic''. This envelope has a peak point $\mu$ that depends on the wave origin $\epsilon$ via eq.~(\ref{eq_mu}). Here, $\epsilon=0.4$ as in Fig.~\ref{fig_comparison} top.}
  \label{fig_model}
\end{figure}

Can we then analytically understand the observed and simulated patterns of shock-induced cavitation? 
Due to axial symmetry, the 3-d geometry of a spherical wave inside a sphere can be reduced to a 2-d circular 
wave reflected inside a circle (Fig.~\ref{fig_model}). Unlike ellipses circles yield no focal point for an eccentric circular wave. Instead, the reflected rays are concentrated within a zone enveloped by the ``catacaustic'' (dashed line in Fig.~\ref{fig_model}), defined as the location where reflected rays intersect. On this catacaustic, we expect a high density of the reflected shock. This model can be extended to 3-d by performing a rotation about the axis of symmetry (horizontal axis in Fig.~\ref{fig_model}). Doing so, we introduce an additional $1/r$-factor in the shock density, where $r$ is the distance from the axis of symmetry. Thus, the location of the highest density of shock-induced cavitation is defined by the intersection of the catacaustic and the axis of symmetry. This intersection lies at a position $\mu$ (see Fig.~\ref{fig_model}) that depends on the wave origin $\epsilon\in[0,1]$. To express $\mu$ as a function of $\epsilon$, consider a single ray emitted at an angle $\alpha\in(0,2\pi)$ and crossing the center line at the position $x$ after its reflection (thick line in Fig.~\ref{fig_model}). From the law-of-sines $\sin(\beta/2)=\epsilon\,\sin\alpha$ and $\sin(\beta/2)=-x\,\sin\gamma$, and trivially $\alpha+\beta+\gamma=\pi$, which solve to
\be
	x = \frac{-\epsilon\,\sin\alpha}{\sin\left[\alpha+2\arcsin(\epsilon\,\sin\alpha)\right]}.
	\label{eq_reflection}
\ee
Eq.~(\ref{eq_reflection}) is meaningful if $x\in[-1,0]$, while otherwise the ray is reflected more than once before crossing the center line. For any $\epsilon\in[0,1]$, the maximum $\mu\equiv\max\{x(\epsilon,\alpha)\in[-1,0]\}$ is reached as $\alpha\rightarrow0$, thus
\be
  \mu=-\epsilon/(2\epsilon+1).\label{eq_mu}
\ee
In conclusion, eq.~(\ref{eq_mu}) specifies the location of strongest shock-induced cavitation. This analytic prediction provides an excellent fit to the microgravity data as well as to the ground-based observations (see dashed crosses and lines in Figs.~\ref{fig_universality} and \ref{fig_comparison}).


\begin{figure*}[t]
  \includegraphics[width=17.2cm]{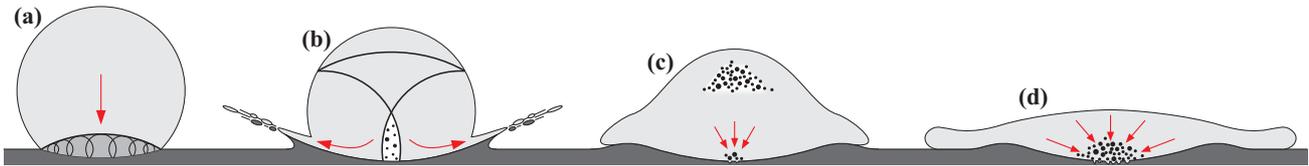}
  \caption{Four mechanism of erosion by an impacting liquid drop (see main text). 
The last mechanism is proposed in this letter. The conditions for these mechanisms differ; hence only some of them may affect a particular impact.}\vspace{-0.2cm}
  \label{fig_impact}
\end{figure*}

Motivated by the observational, numerical, and analytical evidence for shock-induced cavitation in drop and jets, we finally wonder about the potentially erosive implications of this cavitation. In fact, in Fig.~\ref{fig_universality}c the impact of the projectile onto the liquid jet is fast enough for the cavitation bubbles to survive until the projectile reaches them. Since cavitation can erode nearby surfaces \citep{Philipp1998}, this observation suggests that fast liquid--solid collisions may cause erosion via shock-induced cavitation at the far-side of the impact point. To calculate the required impact velocity for a drop with diameter $D$ (e.~g.~a rain drop), we assume that a spherical shock wave is emitted at the initial contact point. Hence, the strongest shock-induced cavitation is located at a distance $2D/3$ from the contact point [eq.~(\ref{eq_mu}) with $\epsilon=1$]. The reflected shock reaches this focus at a time $t_{\rm c}\approx4D/(3c)$ after the impact, where $c$ is the speed of sound, while the impacting solid itself reaches the same point at a time $t_{\rm v}\approx2D/(3v)$, where $v$ is the impact velocity. For erosion to take place, we therefore require $t_{\rm c}+t_{\rm R}\geq t_{\rm v}$, where $t_{\rm R}$ is the life-time of the bubbles \citep{Rayleigh1917}. Hence, a lower impact velocity limit for this type of erosion is\vspace{-0.2cm}
\be
	v\geq\left(\frac{3\,t_{\rm R}}{2\,D}+\frac{2}{c}\right)^{-1}.\label{eq_v0}
\ee
Eq.~(\ref{eq_v0}) is valid for $2v<c$, since otherwise the solid reaches the shock focus before the reflected shock itself. The characteristic bubble life-time $t_{\rm R}$ depends on the liquid and shock parameters, although our cases (Figs.~\ref{fig_comparison}, \ref{fig_universality}b,c) all yield average life times of order $t_{\rm R}=20\rm\,\mu s$ (with the highest values reaching $t_{\rm R}=50\rm\,\mu s$). Adopting $t_{\rm R}=20\rm\,\mu s$ for the case of a typical rain drop ($D=3-4\rm\,mm$, $c\approx1500\rm\,m\,s^{-1}$) eq.~(\ref{eq_v0}) then gives a lower velocity limit of $v\approx100\rm\,m\,s^{-1}$, which is roughly an order of magnitude above the free fall velocity of rain. This reveals that the mechanism of cavitation erosion described in this section will only be active in particular cases, such as aircrafts and missiles eroded by rain \citep{Fyall1966,Zahavi1981} and Pelton turbine blades eroded by high-speed (up to $200\rm\,m\,s^{-1}$) droplets and jets \citep{Perrig2007}.

To complete the picture, Fig.~\ref{fig_impact} presents a synthetic view of the most important mechanisms of erosion that are known to occur during the impact of a liquid drop, including those addressed in the past (damages by such erosion are discussed in Refs.\citep{VanDerZwaag1983,Zahavi1981,Fyall1966}):\\
\emph{(a) Hammer pressure:} On initial contact, erosion can result from the high compression-pressure
 of up to $p=3\rho c v$ for rigid solids \citep{Field1989}, which exists while the contact edge expands supersonically. \\
\emph{(b) Jetting:} When the contact edge becomes subsonic, the pressure-shock detaches from the contact edge and high-speed jets emerge from the latter \citep{Field1989}, causing shear erosion \citep{Springer1976} (e.~g.~crater rims formed by hypersonic impacts of atoms \citep{Samela2008} and meteorites  \citep{Senft2008}).\\
\emph{(c) Near-side cavitation:} A lateral pressure shock travelling from the contact edge to the axis of symmetry can produce cavitation next to the initial contact point \citep{Field1985,Bourne1996}, which may cause point-like erosion \citep{Bourne1995}.\\
\emph{(d) Far-side cavitation:} New mechanism suggested here, which occurs via cavitation caused by reflected shocks opposite the impact point (see also simulations of low pressure at the shock-focus \citep{Haller2002,Sanada2008,Xiong2010}).

We emphasize that a particular drop impact may only involve some of these erosive mechanisms. A systematic study of the applicability of each mechanism as a function of dynamical and geometrical parameters, as well as material properties promises an interesting road for forthcoming research.


This research was supported by the Swiss National Science Foundation (Grants 200020-116641, PBELP2-130895) and the European Space Agency ESA. We thank E.~Robert, M. Rouvinez, and the Swiss Army for contributing to the images in Fig.~\ref{fig_universality}b, c.


\end{document}